\documentclass[osajnl,twocolumn,showpacs,superscriptaddress,10pt]{revtex4-1}
\usepackage{graphicx}

\begin{document}
\title{Ultraslow long-living plasmons with electromagnetically induced transparency}
\author{D. Ziemkiewicz}
\affiliation{Institute of Mathematics and Physics, UTP University of Science and Technology, Al. Kaliskiego 7, 85-789 Bydgoszcz, Poland.}
\author{K. S\l{}owik}
\email{Corresponding author: karolina@fizyka.umk.pl}
\affiliation{Institute of Physics, Faculty of Physics, Astronomy and Informatics, Nicolaus Copernicus University, Grudziadzka 5, 87-100 Torun, Poland}
\author{S. Zieli\'{n}ska-Raczy\'{n}ska}
\affiliation{Institute of Mathematics and Physics, UTP University of Science and Technology, Al. Kaliskiego 7, 85-789 Bydgoszcz, Poland.}




\begin{abstract}
We analytically examine propagation of surface plasmon polaritons (SPPs) at a thin metallic film between glass substrate and electromagnetically-induced-transparency (EIT) medium. 
High-precision and high-resolution in frequency domain provided by EIT paves the way towards plasmonic group velocities reduction even by $4$ orders of magnitude 
and corresponding lifetime enhancement of SPPs up to microseconds. 
\end{abstract}


\maketitle

\section{Introduction}
On-chip devices, based on plasmonics, offer the potential to control and process signals at microscale,
where surface plasmon polaritons are exploited as information carriers.
However, short lifetime of SPPs still belongs to the major bottlenecks in plasmonics.
Circumventing this problem would be a key to applications for nanoscaled signal processing, information transfer or quantum memories. 
So far, the efforts to achieve longer SPP lifetimes were focused on using thin films, where long-range SPP modes can be excited \cite{Berini2009}.

Another challenge in plasmonics relates to novel, controllable ways to influence the propagation velocity of SPPs, to eventually build compact delay lines for plasmons. 
Methods of slowing down SPPs are usually based on structured metallic surfaces \cite{Kocabas2009} or gratings \cite{Sondergaard2006, Chen2008}. None of these ways is really tunable: the plasmonic velocity depends on the geometry of the structure which - once fabricated - can hardly be modified. 
In this work we propose to exploit electromagnetically induced transparency \cite{Harris1997} both for plasmonic slow-down and plasmonic lifetime enhancement.

For this purpose we propose to use thin film geometries, where long-range SPPs can be created. 
SPPs can propagate over considerable distances reaching up to centimeters \cite{Berini2009},
which makes them suitable for processing with external means \cite{Konopsky2006}. 
One of such means takes advantage of the tunability offered by EIT \cite{Yannopapas2009,Du2012,Shen2014,Ziemkiewicz2017}. 
EIT has already been proposed for slowing down SPPs near the interference between dielectric and active, negative index metamaterial \cite{Kamli2008}.

\begin{figure}[ht!]
\centering
\fbox{\includegraphics[width=\linewidth]{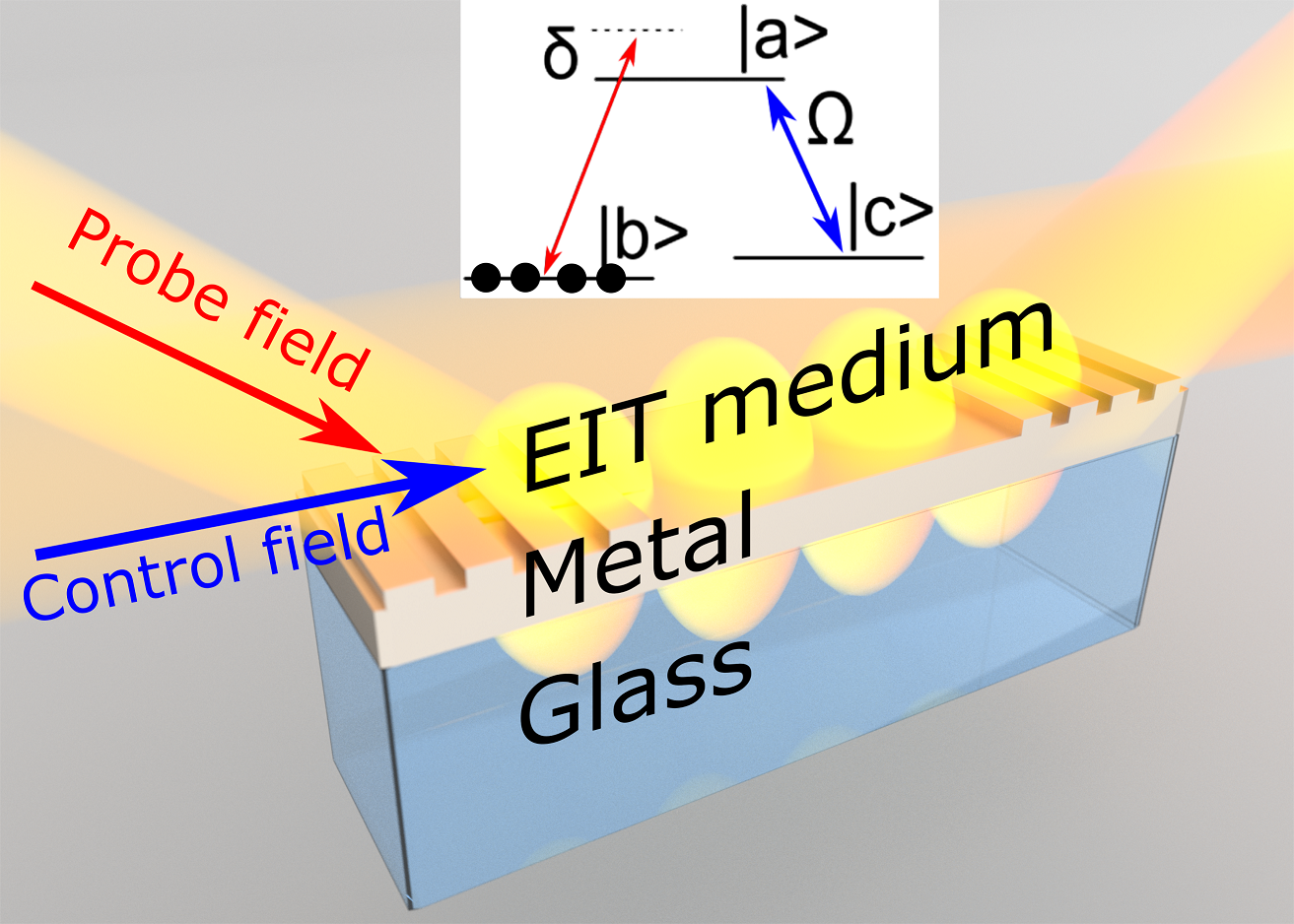}}
\caption{A probe beam incident at a silver film between a glass surface and EIT medium excites propagating surface plasmon polaritons.
The inset shows an atomic $\Lambda$ system enabling realization of the EIT phenomenon.}
\label{fig:setup}
\end{figure}
In the EIT phenomenon, dispersive properties of a medium can be externally tuned through illumination with a moderately strong laser beam, referred to as a control one. 
With such illumination it is possible to almost cancel absorption of a weaker resonant probe. 
As a result the probe, instead of being efficiently damped, can propagate at a reduced group velocity which depends on the control field's intensity \cite{Harris1997, Fleischhauer2005}. 

In this work we propose to make use of such tunable dispersive properties to reduce the velocity of SPPs and significantly increase their lifetime. 
An exemplary setup where such slowdown could be realized consists of a silver thin film on top of a glass substrate, with coupling and decoupling gratings (Fig.~\ref{fig:setup}).
The probe beam is incident at an angle $\theta$ chosen such that it excites a pair of SPPs, propagating along the thin film. 
This pair consists of a long-range and short-range SPPs (LRSPP and SRSPP, respectively), characterized by different group velocities and propagation distances \cite{Berini2009}. 
Instead of embedding the setup in air or a dielectric, we suggest to put it in a tunable EIT medium, 
with the control beam co-linear with the propagation path of the SPPs along the film. 
We show analytically and verify numerically that using EIT could enable slow, but long-living plasmons. 

\section{Electromagnetically induced transparency}
Before we proceed to discuss the details of the scheme, we introduce the basics of the EIT phenomenon.

\begin{figure}[htbp]
\centering
\fbox{\includegraphics[width=0.5\linewidth]{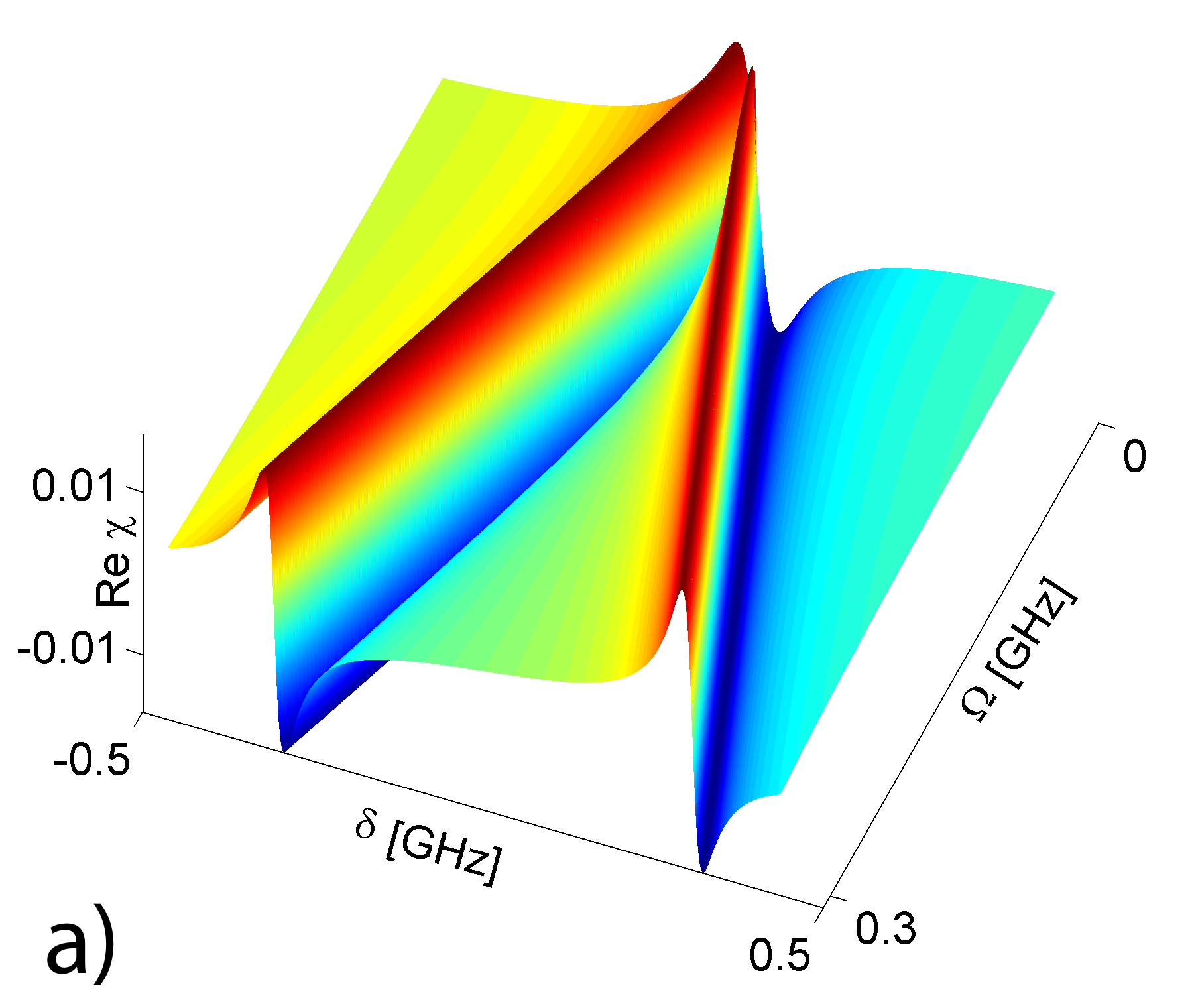}\includegraphics[width=0.5\linewidth]{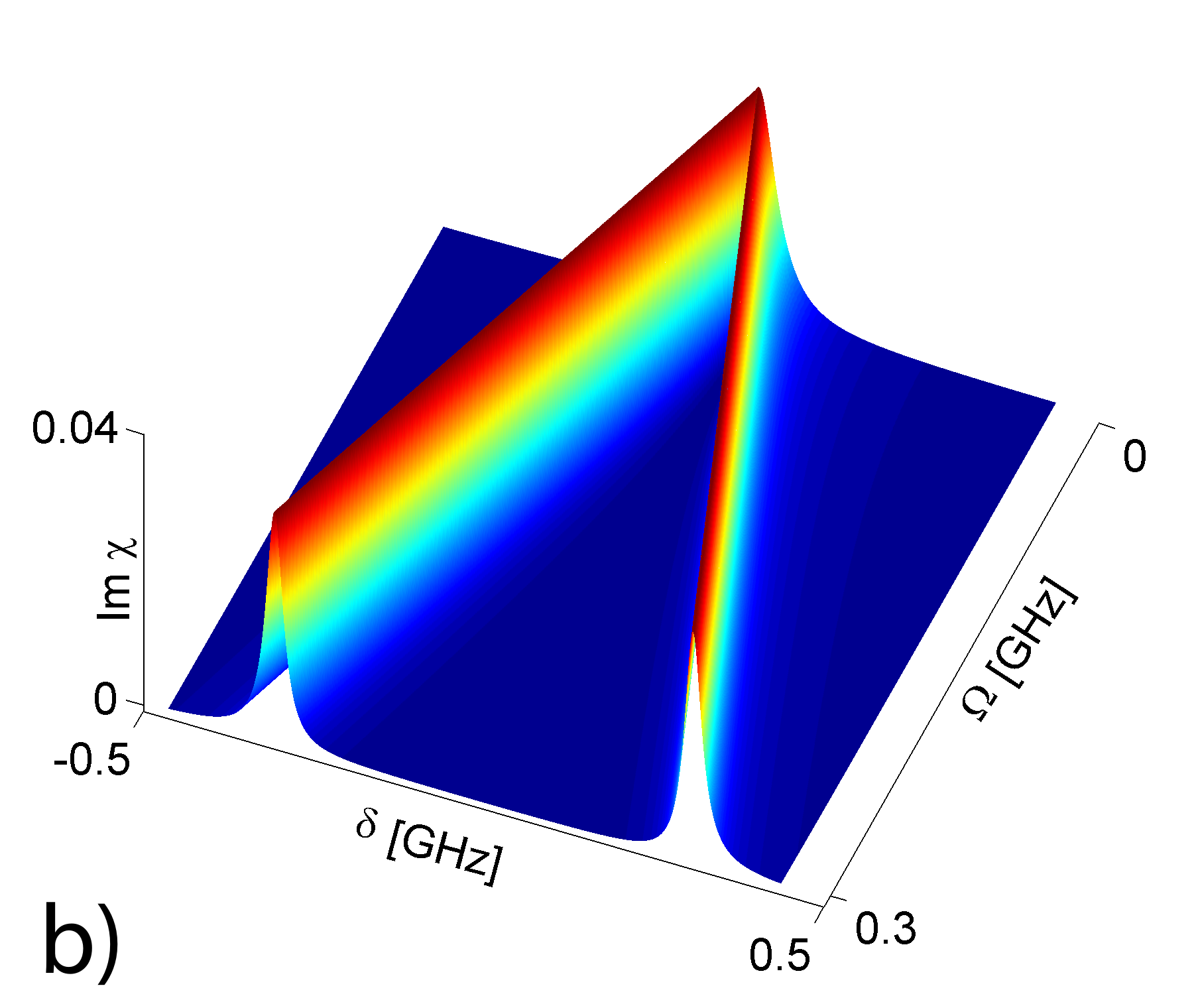}}
\caption{Real (a) and imaginary (b) part of the electric susceptibility of the EIT medium in function of probe detuning $\delta$ and control field $\Omega$.}
\label{fig:chi}
\end{figure}

The simplest EIT can be realized in an atomic medium whose energy structure can be approximated by the three-level $\Lambda$ configuration with the excited state $a$, fully populated ground state $b$ and a side state $c$ (inset of Fig.~\ref{fig:setup}). 
The transition frequencies between the states $i$ and $j$ are denoted as $\omega_{ij}$.
The control beam drives the $c \leftrightarrow a$ transition, while the probe couples $b$ and $a$ levels.
Once the control beam coupling two empty states creates coherence in the medium, the probe beam can propagate at a reduced velocity and almost without absorption \cite{Fleischhauer2005}. 

A quantitative description of the modified dispersive response is given through
the electric susceptibility $\chi(\delta,\Omega)$ of the $\Lambda$ medium
\begin{equation} \label{eq:chi}
 \chi(\delta,\Omega) = -\frac{N|\mu_{ab}|^2}{\hbar\epsilon_0}\frac{\delta-\delta_{c}+i\gamma_{bc}}{(\delta+i\gamma_{ab})(\delta-\delta_{c}+i\gamma_{bc})-|\Omega|^2}, 
\end{equation}
where $\delta = \omega-\omega_{ab}$ and $\delta_c=\omega_c-\omega_{ac}$ represent the detuning from the atomic transition resonance of the probe and control beams, centered at $\omega$ and $\omega_c$, respectively.
The symbol $\gamma_{ij}$ stands for the decoherence rate at the $i \leftrightarrow j$ transition, $N$ is the atomic density, $\mu_{ab}$ is the transition dipole moment, 
$\hbar$ is the reduced Planck's constant, $\Omega$  is related to the control field $\mathbf{E}_c$ by \mbox{$\Omega=\frac{\mathbf{E}_c.\mu_{ac}}{\hbar}$}, and $\epsilon_0$ stands for the vacuum electric permittivity. 
The imaginary part of the susceptibility [Fig.~\ref{fig:chi}(b)] shows a prominent dip centered around the atomic resonance, referred to as the transparency window. 
Its width can be externally tuned: it is proportional to the control field $\Omega$. 
The dispersion, i.e. susceptibility's real part [Fig.~\ref{fig:chi}(a)] is directly responsible for the probes group velocity
\begin{equation}
v_g(\omega,\Omega) =\frac{c}{n(\omega,\Omega)+\omega \frac{d n}{d \omega}}, \label{eq:velocity}
\end{equation}
where in this case $n(\omega,\Omega)=\sqrt{1+\Re[\chi(\omega,\Omega)]}$, but it will be modified at presence of the metallic film. Please note that the group velocity can be tuneded via the intensity of the control field.
 The dispersion inside transparency window becomes normal, linear to a good approximation and $Re[\chi(\omega,\Omega)]\sim \frac{1}{\Omega^2}$, with slope incerases for decreasing control field \cite{Fleischhauer2005}. This means that the probe pulse travels  with the reduced group velocity (with respect to the vacuum speed of light $c$) leading to its slowdown  even by several orders of magnitude \cite{Hau1999, Marangos1999}, or a possible complete stop \cite{Phillips2001,Liu2001}.

\section{plasmonic slowdown}
We now proceed to discuss the details of the scenario proposed in this work. 
Adjusting the thickness of the silver layer and the incidence angle, it is possible to excite SPPs propagating along the metallic film. 
This is achieved with the TM-polarized probe beam illuminating the setup.
In case of such a thin metal layer with thickness $d$ much smaller than the optical wavelength $d << \lambda$, the probe creates two distinct SPP modes, called long-range- and short-range surface plasmon polaritons. 
As compared to SPPs on a thick metal layer, the described plasmons, especially the long range modes, are characterized by a very long, nanosecond lifetimes 
and propagation distances on the order of millimetres \cite{Konopsky2006,Yi2007,Park2009,Berini2009}. This key characteristic of thin film plasmons is crucial in practical realization of our scheme, 
allowing for macroscopic sample size, as compared to other schemes based on a typical Kretschmann configuration \cite{Kamli2008,Shen2014}.

EIT conditions are created with a TE polarized control beam parallel to the metallic film (see Fig.~\ref{fig:setup}). 
In the situation when the $\Lambda$ system is constituted by two hyperfine-splitted metastable states $b$ and $c$, the Doppler-broadening has no adverse effect on EIT even at room temperatures, provided that one uses co-propagating probe and control beams \cite{Harris1997, Alzetta2004,Fleischhauer2005}.
The role of the control beam is solely to modify the optical properties of the EIT medium, i.e. to change the propagating conditions for the probe SPPs. 
At presence of the control field $\Omega$ the dispersive properties of the EIT medium are modified as described above  and given by Eq.(\ref{eq:chi}). 
The resulting dispersive properties of the probe SPP modes follow from the boundary conditions on the glass-metal and EIT-metal interfaces \cite{Economou1969,Raether1988,Pitarke2006}
\begin{eqnarray}\label{eq:disp}
\frac{\kappa_m\tanh(\kappa_m\frac{d}{2})}{\epsilon_m} = \frac{-\kappa_{eit}}{\epsilon_{eit}} = \frac{-\kappa_g}{\epsilon_g},\nonumber\\
\frac{\kappa_m\coth(\kappa_m\frac{d}{2})}{\epsilon_m} = \frac{-\kappa_{eit}}{\epsilon_{eit}} = \frac{-\kappa_g}{\epsilon_g},
\end{eqnarray}
where $\kappa_j = \sqrt{k^2 - \epsilon_j\frac{\omega^2}{c^2}}$ is the component of the wave vector $k$ parallel to the surface in the $j$-th medium with $j\in\{eit,m,g\}$ corresponding respectively to the EIT medium, metal and glass, 
$\epsilon_j$ are relative electric permittivities, and \mbox{$\epsilon_{eit}(\omega,\Omega) = 1+\chi(\omega,\Omega)$}.

For a quantitative discussion, we consider the following parameters: 
a probe field with $\lambda = 589$ nm exciting a pair of SPPs on a silver layer ($\epsilon_m = -13.3 + 0.883 i$) \cite{Palik1998} 
with thickness $d=32.7$ nm. 
The metal film is surrounded by glass of $\epsilon_g = 2.2$ and EIT medium.
The latter consists of sodium vapour, with typical densities in the range of $N=10^{10} - 10^{12}~\mathrm{cm}^{-3}$,
illuminated with the control field $\Omega$ typically of several hundred MHz. 
The chosen wavelength $\lambda$ corresponds to the sodium D2 line ($3^{2}S_{1/2}\rightarrow 3^{2}P_{3/2}$).
The upper state $a$ would simply be the $|3^{2}P_{3/2}, F=0, m_F=0\rangle$ hyperfine sublevel, while
the lower states would correspond to symmetric and antisymmetric superpositions, respectively:
$b,c \sim |3^{2}S_{1/2}, F=1,m_F = -1\rangle \pm |3^{2}S_{1/2}, F=1,m_F = +1\rangle$ \cite{Steck}.
This is necessary for them to be coupled with linearly polarized light, as required by our scenario:
the $c \leftrightarrow a$ transition is driven with TE-polarized light (the control field),
while the probe of TM polarization couples states $b$ and $a$ [Fig.~\ref{fig:setup}].
We emphasize that such EIT scenario has been demonstrated experimentally for sodium vapours at room temperatures \cite{Alzetta2004}.

The electric dipole moment of the $a \leftrightarrow b$ transition is $\mu_{ab} = 1.72 \times 10^{-29}$ Cm. 
The most important decoherence mechanism in sodium is spontaneous emission $\gamma_{ab, ac} = 60$ MHz, whose influence is much greater than the impact of collisions with cell walls. 
These collisions can be further reduced with an addition of buffer gas, which is a standard procedure in EIT experiments. 
On the contrary, dephasing $\gamma_{bc}$ between the lower states corresponds to an electric-dipole-forbidden transition, for which spontaneous emission is negligible, so this decoherence rate is small and determined by atomic movement. Only the atoms coherently prepared by the control beam contribute to EIT. In the heated cell the time an atom spends on average within the volume illuminated by the control beam can be estimated as  $T=15$ $\mu$s \cite{Dziczek2005}. An additional depolarizing process is related to collisions \cite{Dziczek2005} with the metallic film that may occur at time-scales of $T^\prime = 0.1$ ms. The total depolarizing rate  $\gamma_{bc}=T^{-1}+{T^\prime}^{-1} \approx 68$ kHz  is small with respect to both: other decoherence processes in the EIT medium and loss in metals. 
Finally, please note that all the important features of EIT remain observable even when $\gamma_{bc} \neq 0$, provided that the control field satisfies $|\Omega|^2 \gg \gamma_{ab} \gamma_{bc}$ \cite{Fleischhauer2005}, which is relatively easily achieved. The width of the transparency window is limited by the Stark shift of atomic energy levels, which happens around \hbox{$\Omega > \gamma_{ab} \approx 500$ MHz}.

\begin{figure}[htbp]
\centering
\fbox{\includegraphics[width=\linewidth]{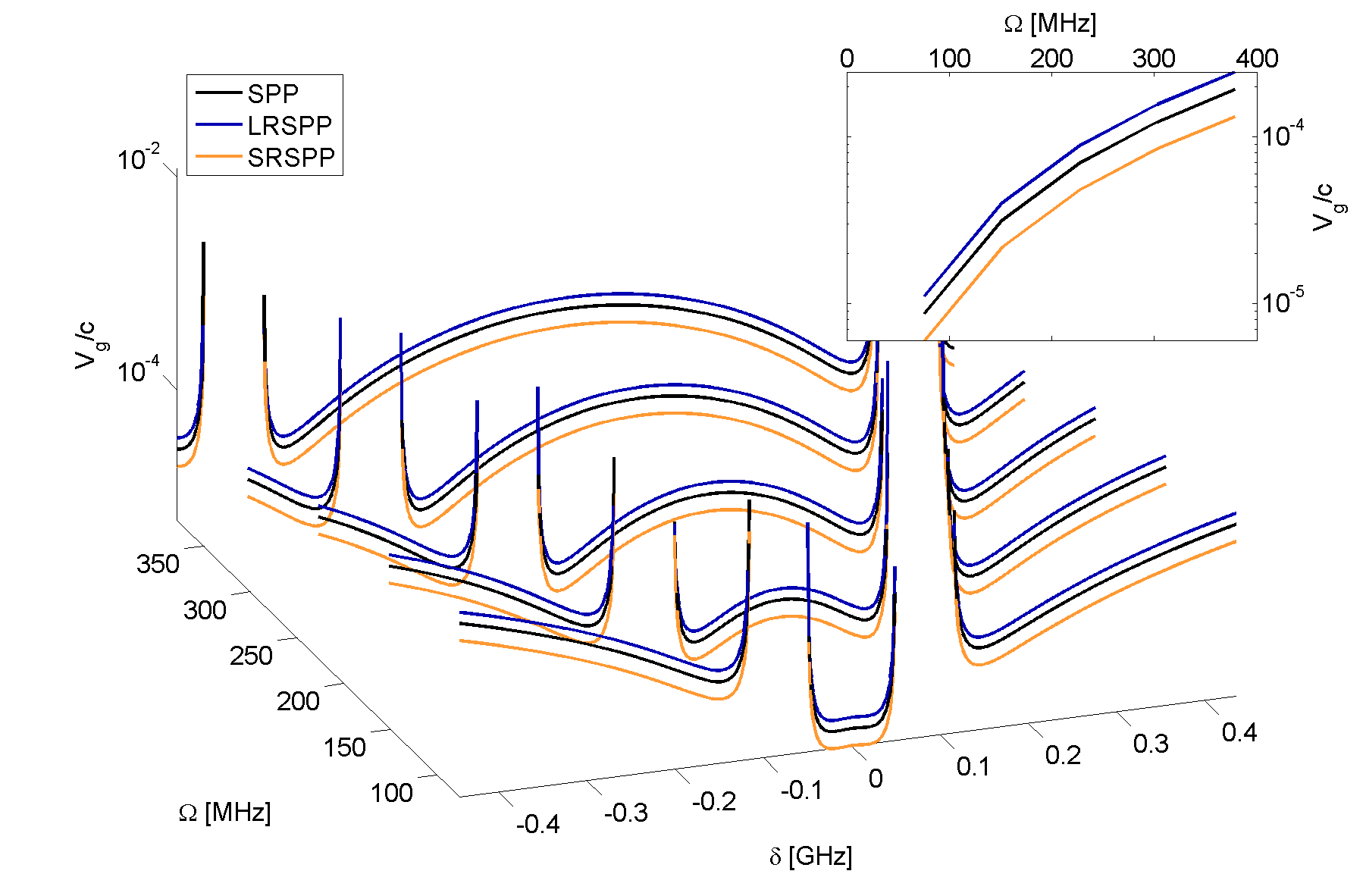}}
\caption{Group velocities of SPPs as a function of the probe detuning $\delta$ and control field $\Omega$. 
The inset shows a cut for $\delta=0$.}
\label{fig:group_velocity}
\end{figure}

Solving Eqs.~(\ref{eq:disp}) numerically, one can obtain the group velocity $v_g = \frac{d \omega}{d k}$ as a function of tunable parameters such as the control field $\Omega$ and the density of the EIT medium $N$.
We plot the resulting group velocity for the set of data given above as a function of the probe detuning $\delta$  from the $b \leftrightarrow a$ atomic transition and the control field $\Omega$ (Fig.~\ref{fig:group_velocity}).
Naturally, LRSPPs (SRSPPs) are characterized with group velocities enhanced (decreased) with respect to SPPs in standard Kretschmann configuration (black line). 
The influence of the EIT medium can be recognized from the characteristic pattern corresponding to the transparency window in Fig.~\ref{fig:chi}(b). 
For frequencies around $\delta=0$, a reduction by the factor of $10^{-4}$ with respect to vacuum speed of light is obtained (see inset). 
The velocity can be tuned by optical means, i.e. by adjustment of the electric field of the control beam $\Omega$. 

\begin{figure}[htbp]
\centering
\fbox{\includegraphics[width=\linewidth]{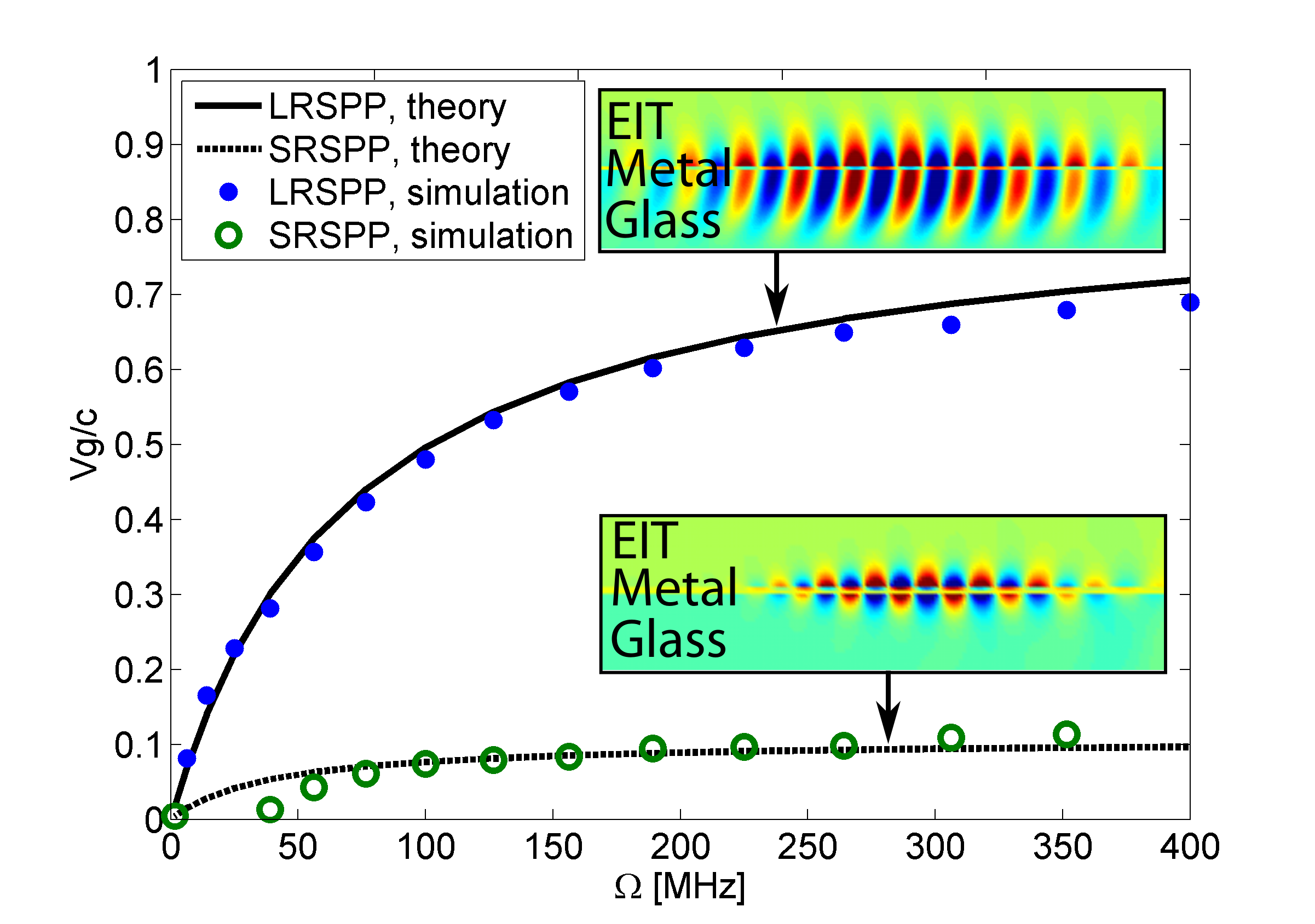}}
\caption{Group velocity of SPPs obtained analytically by solving Eq.~(\ref{eq:disp}) and numerically using FDTD.
The insets show distributions of the electric field component perpendicular to the metallic surface for LR- and SRSPP modes.}
\label{fig:numeryka}
\end{figure}

Our analytical findings are verified by a direct comparison with numerically obtained results for group velocity reduction (Fig.~\ref{fig:numeryka}), where excellent agreement was achieved. 
We have used the finite-difference time-domain (FDTD) method, based directly on Maxwell's equations in 2D complemented with suitable dispersion model for metal and EIT medium \cite{Ziemkiewicz2015,Ziemkiewicz2017}. 
Please note that the asymptotic value of the group velocity obtained for large control fields $v_g \approx 0.8632~c$ would be the SPP velocity at the absence of the EIT medium. 
Due to simulation time and stability constraints, we numerically investigated the case of very low atomic medium densities $N = 10^9$ cm$^{-3}$, resulting in a slowdown by up to two orders of magnitude. 
As one can see from Eqs.(\ref{eq:chi}-\ref{eq:velocity}), since $\chi \sim N$, the results are approximately scaled by the factor $1/\sqrt{N}$, which allows us to anticipate smaller group velocities for enhanced densities $N$, 
in accordance with the analytical findings. 
Application of the numerical method enables an illustration of field distributions corresponding to propagation of LR- and SRSPPs along the thin film (see insets). 
Please note that a considerable fraction of the field enters the EIT medium. 

\begin{figure}[htbp]
\centering
\fbox{a)\includegraphics[width=0.48\linewidth]{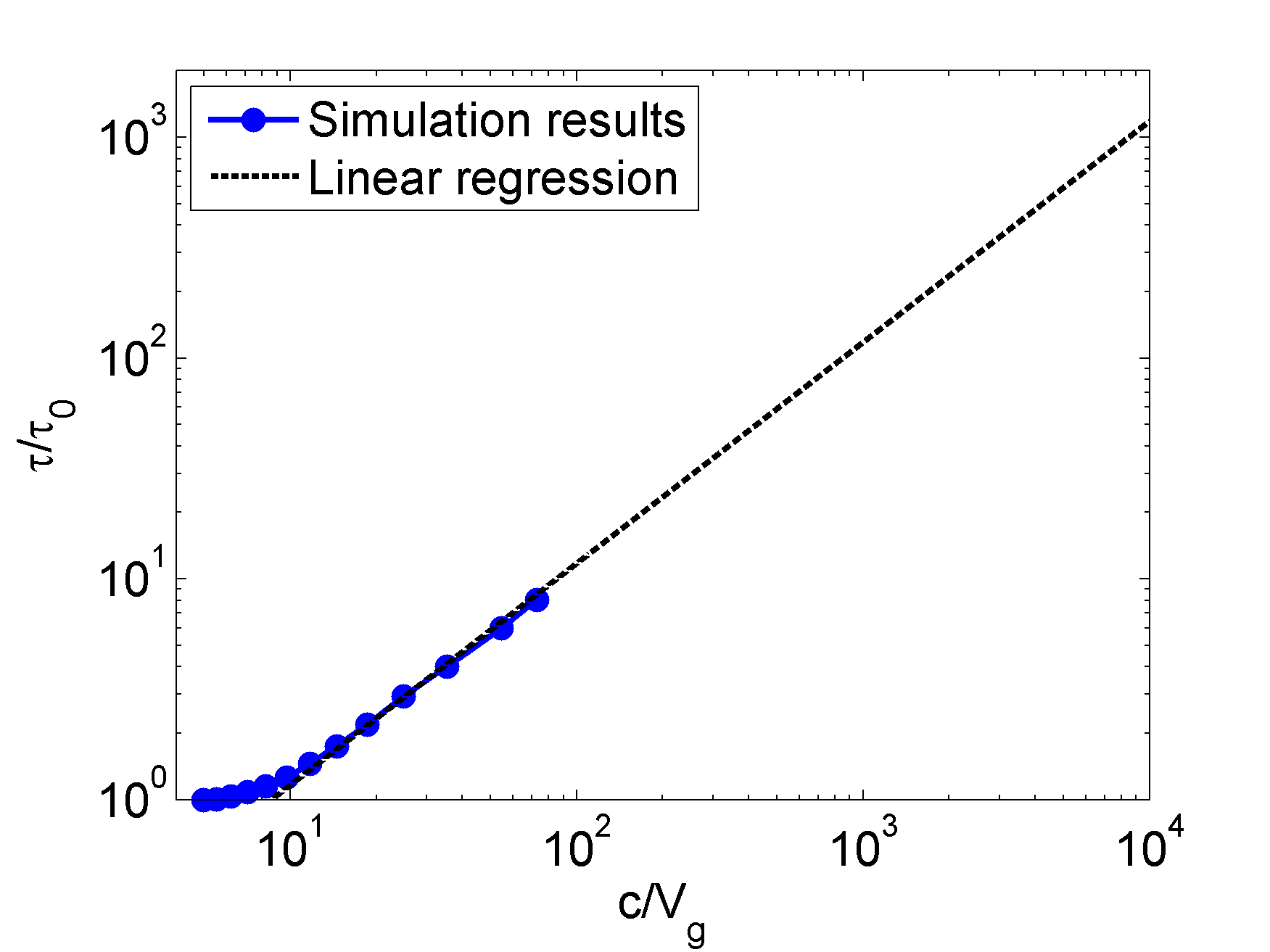}b)\includegraphics[width=0.48\linewidth]{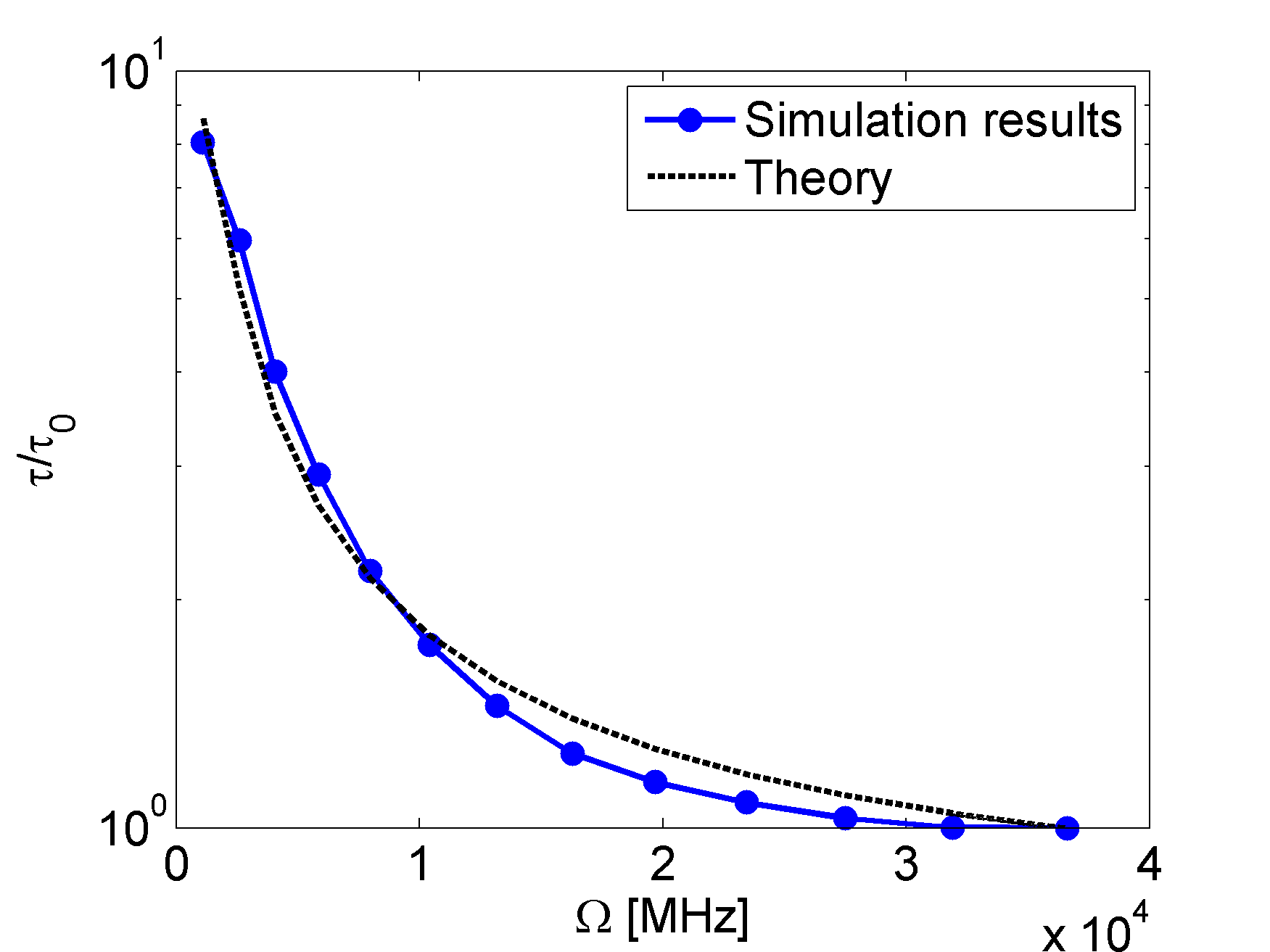}}
\caption{LRSPP lifetime as a function of $v_g$ (a) and $\Omega$ (b). The linear regression plot in (a) was fit to the points corresponding to $v_g<0.1c$. }
\label{fig:tau}
\end{figure}

The reduced group velocity is directly related to a strongly enhanced plasmonic lifetime. 
For strong velocity reduction $v_g\ll c$ we have $\epsilon + \omega\frac{\partial \epsilon}{\partial \omega} \gg 1$, and obtain 
\begin{equation}\label{eq:2c}
 v_g \approx \frac{2c}{\epsilon + \omega\frac{\partial \epsilon}{\partial \omega} }.
\end{equation}
The probe beam energy density depends on its electric field $E$ \cite{Landau1960}
\begin{equation}\label{eq:w}
W = \frac{\partial (\omega \epsilon)}{\partial \omega}|E|^2 = (\epsilon + \omega\frac{\partial \epsilon}{\partial \omega} ) |E|^2 \approx \frac{2c}{v_g} |E|^2.  
\end{equation}
For SPP in form of an impulse taking finite time $\delta t$, the volume $V$ taken by an SPP is proportional to its group velocity, e. g. $V \sim v_g \delta t$. Therefore, the total field energy $\int_V W d^3x$ is independent on $v_g$. 
The power absorbed in the metal, which is the dominant source of losses in the system, is expressed as
\begin{equation}\label{eq:P}
 P = \int_V \sigma |E_{m}|^2 d^3x \sim \sigma \frac{v_g }{2c} \int_V W d^3x \sim v_g,
\end{equation}
where $\sigma$ is the conductivity of the metal, $E_m \sim E$ is the field inside the metal, and we have used Eq.~(\ref{eq:w}). 
Therefore, the absorbed power is proportional to the group velocity, and can be greatly reduced in conditions of EIT. 
In other words, longer lifetimes origin at the fact that a reduced group velocity in EIT conditions corresponds to energy transfer from the SPP field to atomic excitations.
This prevents the metal from absorbing the field. 
This is the reason for the increased plasmonic lifetimes $\tau$, compared in Fig.~\ref{fig:tau}(a) to lifetimes $\tau_0$ at the absence of the EIT medium.
As expected, we see an almost linear dependence of lifetime on the inverse of the group velocity for significant slowdowns. 
The results are extrapolated to the regime of smaller group velocities achieved for atomic densities typically used in EIT experiments. The obtained correlation between SPP lifetime and group velocity agrees with the findings of Derrien \emph{et al} \cite{Derrien2016}, where moderate lifetime enhancement has been discussed at various metal-air interfaces. It should be stressed that the increased lifetime is crucial for the effective slowdown of SPPs; the slowdown by two orders of magnitude results in $\tau \approx 1$~ns and corresponding reduced plasmon spectral width \hbox{$\Gamma \approx 1$~GHz}, which is sufficiently narrow to fit inside the transparency window.

We have examined the same lifetime enhancement in dependence of the control field $\Omega$ [Fig.~\ref{fig:tau}(b)], illustrating the optical tunability of the SPP lifetime.
Decreasing the control field leads to a narrower transparency window, which results in lower group velocities and longer lifetimes. The numerical results are compared to theoretical relation given by Eq.~(\ref{eq:w}) and Eq.~(\ref{eq:P}), where the approximation (\ref{eq:2c}) has not been used. In such a case, one obtains nonlinear relation between SPP lifetime and control field 
\begin{equation}
\tau \sim 1 + \chi(\omega,\Omega) + \omega\frac{\partial \chi(\omega,\Omega)}{\partial \omega}
\end{equation}
which is a fairly good match to the numerical results in a whole group velocity range.

\section{Conclusions}
We have proposed to exploit the EIT phenomenon for a significant reduction of plasmonic propagation velocity and the corresponding lifetime enhancement.
Both group velocity and lifetime can be optically tuned with changes of the control field.
We have discussed an exemplary setup feasible to realize our predictions: a silver thin film placed on a glass substrate and surrounded by sodium vapours, where the EIT is performed. 
This finds potential applications for tunable technologies of nanoscaled information processing both at a classical and at a quantum level, nanodevices that could be switched between their operational modes, or even quantum memories for plasmons.
\bigskip



\bibliography{eit}

\end{document}